\documentclass[aps,floatfix,showpacs,twocolumn,prl]{revtex4}
\usepackage[dvips]{graphicx}
\usepackage{amsfonts}
\usepackage{amssymb}\usepackage{times}
\usepackage{sub_JP}
\usepackage{MHequ}

\def\L{{\rm L}}
\def\R{{\rm R}}

\def\eref#1{(\ref{#1})}
\let\rho=\varrho

\begin{document}
\title{Magnetically Induced Thermal Rectification}
\author{Giulio Casati${}^{1,2,3}$}
\author{Carlos~Mej\'\i{a}-Monasterio${}^4$}
\author{Toma\v z Prosen${}^5$}
\affiliation{Center for  Nonlinear and  Complex  Systems, Universit\`a
  degli Studi dell'Insubria, Como Italy${}^1$}
\affiliation{CNR-INFM and Istituto Nazionale di Fisica Nucleare, Sezione di Milano${}^2$}
\affiliation{Department of Physics, National University of Singapore, Republic of Singapore${}^3$}
\affiliation{Dipartimento di Matematica, Politecnico di Torino, Italy${}^4$}
\affiliation{Physics Department,  Faculty of Mathematics  and Physics, University  of Ljubljana, Ljubljana, Slovenia${}^5$}
\date{\today}

\begin{abstract}
  We consider far from equilibrium heat transport in chaotic billiard chains
  with non-interacting charged particles in the presence of non-uniform
  transverse magnetic field.  If half of the chain is placed in a strong
  magnetic field, or if the strength of the magnetic field has a large
  gradient along the chain, heat current is shown to be asymmetric with
  respect to exchange of the temperatures of the heat baths.  Thermal
  rectification factor can be arbitrarily large for sufficiently small
  temperature of one of the baths.
\end{abstract}

\maketitle

The problem of explaining irreversible macroscopic transport laws, such as the
Fourier law of heat conduction, from the reversible microscopic equations of
motion is one of the open problems in nonequilibrium statistical mechanics
\cite{bonetto,lepri}.  The problem is still far from being settled, in
particular in the framework of low dimensional systems, e.g. one dimensional
particle chains, where different instances of anomalous transport can be
identified due to various kinematic or dynamic mechanisms.

In systems of non-interacting particles, e.g. quasi 1-d chaotic billiards, the
validity of Fourier law has clearly been confirmed \cite{alonso}, even though
such non-interacting system cannot exhibit local thermal equilibrium in the
non-equilibrium steady state \cite{dhar,mejia}. Furthermore, interesting
connection between the Fourier law and particle diffusion has been established
\cite{liwang}.  These problems are not only interesting for understanding the
fundamentals of statistical mechanics, but they may also have straightforward
applications, e.g.  for connecting dynamics and transport in emerging
nano-technology, engineering of molecular motors, understanding and control of
energy flow in bio-molecules, etc.

In view of these ideas, Terraneo et al. \cite{terraneo} have recently proposed
a  mechanism  for  thermal  rectification  in  an  anharmonically  interacting
particle chain.  Using an  effective phonon approach,  they have shown  that a
chain composed  of three  different parts may  have asymmetric  heat transport
properties  due to  (non)matching  of  the effective  phonon  bands. This  and
related  ideas have  been further  elaborated and  improved\cite{li-2,li-3}. A
further step to devise a  thermal transistor has been discussed in \cite{li-4}
in  terms of  the negative  differential thermal  resistance observed  in some
anharmonic chains.

Since  these  first  works,  other  different mechanisms  leading  to  thermal
rectification  have been  described. In  \cite{segal}  it was  shown that  the
nonlinearity of  the dynamics  of an asymmetric  two-level system leads  to an
asymmetric heat flow. More recently  in \cite{EMM-2} it was shown that thermal
rectification  can   be  observed  in  asymmetric   billiards  of  interacting
particles.  When the billiard is subjected to an external temperature gradient
the  effective   interaction  leads  to   a  temperature  dependence   of  the
transmission coefficient and thus, it  is possible to dynamically controls the
transmission probability of  the billiard.  For this type  of billiard systems
rectifications as large as $10^3$ were observed. However, the theory presented
in  \cite{EMM-2}  requires  the  knowledge  of  the  microscopic  transmission
coefficients,   which   is,  at   best   phenomenological.   Another,   simple
phenomenological   mechanism  of  thermal   rectification,  based   on  strong
positional and temperature dependence  of local thermal conductivity, has been
proposed by Peyrard \cite{peyrard}. 

A common problem with most of these proposals is that the rectification factor
has  always been  rather limited,  i.e.   it is  very difficult  to achieve  a
situation in  which heat flows only in  one direction and not  in the opposite
one. Moreover,  the fact  that the thermal  rectification depends  directly or
indirectly on  the microscopic particle-particle interaction,  renders difficult the  ability to
control the power of rectification.

In  this  letter  we  propose   a  novel  microscopic  mechanism  for  thermal
rectification which  works in the  absence of particle interactions. Instead,
thermal  rectification is  controlled  with an  external non-uniform  magnetic
field, leading  to an {\em arbitrarily  large} power of  rectification.  In our
model heat  is carried by charged  particles and the only  restriction is that
the  typical mean-free-paths  due  to dissipative  mechanisms  should be  much
larger than the Larmor radius of  the charged particles in the magnetic field. 
In practice this means that the temperatures of the baths should be quite low,
but  such regimes  are nowadays  easily accessible  for example  in mesoscopic
physics,  quantum   dots/wires,  anti-dot  lattices   etc.  The  rectification
mechanism is  very simple, it  is based on  asymmetric reflection of  slow and
fast particles off the interface  between regions with different magnetic field
intensities and  predicts  arbitrary  large  rectification factors  for  sufficiently  low
temperatures.

\begin{figure}[!t]
\begin{center}
  \includegraphics[scale=0.7]{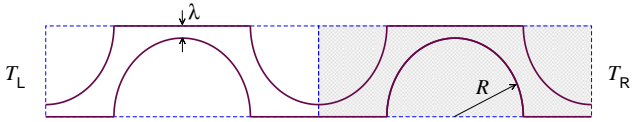}
\caption{
  Geometry of the model: We consider a billiard composed by $N$ symmetric unit
  cells, each of which is made of a rectangular boundary of width $\delta y   2/\sqrt{3}$ and length $\delta x = 4$. Inside each cell there are three
  circulars hard scatterers of radius $R$ disposed in a triangular lattice
  (see the text).  Each cell is subjected to an external perpendicular uniform
  magnetic field of strength $B_i \in [0,B]$. At the left and right boundaries
  the billiard is placed in contact with two stochastic thermal baths at
  different temperatures $T_\L$ and $T_\R$.  The figure corresponds to a
  channel of $N=2$ cells in which $B_1=0$ and $B_2=B$.  The dashed lines are
  drawn as a reference.
\label{fig:channel}}
\end{center}
\end{figure}

We consider a gas of noninteracting point particles of mass $m$ and electric
charge $e$ that move freely inside a two-dimensional billiard region. The
billiard is shaped as a chain of equal cells as depicted in
Fig.~\ref{fig:channel}. Let the circular obstacles have radius $R$ and centers
of the discs be arranged in a hexagonal array with lattice distance
$4/\sqrt{3}$, such that the billiard motion has a closed horizon for $R \ge
1$.  Then we cut out a quasi one-dimensional billiard channel of width $\delta
y=2/\sqrt{3}$ through the centers of two nearby rows of discs, so that one
rectangular $\delta x \times \delta y$ cell of the channel, containing one
half disk and two quarter disk obstacles, has length $\delta x=4$.  Negative
curvature of the billiard boundary ensures that the motion in the absence of
the magnetic field is completely chaotic - hyperbolic.

The simplest model that we consider is composed of two cells
(Fig.~\ref{fig:channel}).  The left cell contains no magnetic field, whereas the
right cell is subjected to a perpendicular uniform magnetic field of strength
$B$. In what follows we will refer as \emph{step} configuration, to a channel
of $N$ cells for which the $N/2$ left cells contain no magnetic field and the
$N/2$ right cells contain a magnetic field of density $B$.  Let
$\lambda$ denote the smallest length scale in the problem, namely the width of
the opening between the neighboring cells in our model,
\begin{equ} \label{eq:lambda}
\lambda = \frac{2}{\sqrt{3}} - R ~.
\end{equ}

The transmission probability of the billiard is controlled by the strength of
the magnetic field. Consider the particles that cross the junction between the
two cells from \emph{left to right}.  There exist a critical velocity $v_c$:
fast particles of velocity $v > v_c$, always enter the right cell, and thus
contribute to the left to right energy flow provided they reach the right end
of the system which is coupled to a heat bath as explained below.
Instead, slow particles of velocity $v < v_c$, such that the
gyro-magnetic radius $\rho(v) = mv/(eB)$ is less than $\lambda/2$ will be
reflected or transmitted depending on the position at which they reach the
interface.

Using a statistical ensemble of trajectories, the condition for the critical velocity 
$\rho(v_c) = \lambda/2$ can be rewritten as the condition giving a critical
temperature
\begin{equ} \label{eq:Tc}
T_{\mathrm{c}} = \frac{(eB_{\mathrm{c}} \lambda)^2}{8m k_{\rm
    B}} ~.
\end{equ}
such that particles which are colder than $T_{\rm c}$ will be reflected in
their majority.

However, for the particles that cross the junction from \emph{right to left}
there is no condition on their velocity and they always enter the left cell. The
above qualitative argument makes it clear that the transport of heat will be
strongly asymmetric with respect to exchange of effective temperatures of
particles on different sides of magnetic field boundary, provided the
temperatures are strongly different, one being larger and the other smaller
than $T_{\rm c}$.  In the rest of the letter we measure the temperature in units
of $T_{\rm c}$, and denote it as $\tau = T/T_{\rm c} = \frac{m}{2 k_{\rm B}
  T_{\rm c}} \langle v^2 \rangle$.

We  couple  the  left  cell  with  a stochastic  heat  bath  \cite{alonso}  of
dimensionless temperature $\tau_{\rm  L}$ and the right cell  with a heat bath
of temperature  $\tau_{\rm R}$, namely if  a particle reaches  the end boundary of
the  billiard that  is  in contact  with a  heat  bath it  is reflected  with a
velocity chosen from a distribution with probability densities
\begin{equa}[3] \label{dist}
&P_n(v_x) & \ \ = \ \ & \frac{1}{\tau}|v_x|\exp\left(-\frac{v_x^2}{2\tau}\right) \ ,\\
&P_t(v_y) & \ \ = \ \ &  \frac{1}{\sqrt{2\pi\tau}}\exp\left(-\frac{v_y^2}{2\tau}\right) \ ,
\end{equa}
where $\tau$ is  the temperature of the respective  heat bath in dimensionless
units.

\begin{figure}[!t]
\begin{center}
\includegraphics[scale=0.65]{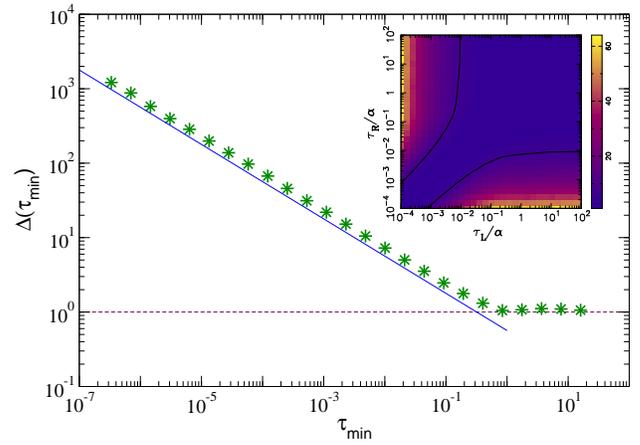}
\caption{(Color online).
  Rectification index $\Delta$ as a function of the minimal temperature
  $\tau_{\min}$ for a $2$-cell channel in a \emph{step} configuration. The
  maximal temperature was set to $\tau_{\max}=33.4275$. The dashed line is for
  the no rectification value $\Delta=1$.  The solid line corresponds to
  $\tau_{\min}^{-1/2}$. Inset: Color density plot for $\Delta$ as a function
  of the temperatures of both baths.  Note that by construction the matrix is
  symmetric. For the sake of presentation the temperatures are rescaled by
  $\alpha=1.33711$.
\label{fig:rect-2}}
\end{center}
\end{figure}

If the left bath is cold, i.e.  $\tau_{\rm L} \ll 1$, then the particles will
most of the time remain in the left cell and there will be no heat current
between the baths, irrespective of the temperature of the right bath, which we
assume is larger than $1$.  If we exchange the temperatures of the baths, then
we will in general have some non-small heat current flowing, since cold
particles have no problem in leaving the region with a magnetic field.

Let us measure the heat currents per particle in the steady state, and denote
them as $J^+$ if $\tau_{\rm L} < \tau_{\rm R}$ and $J^{-}$ if the temperatures
are exchanged, i.e. $\tau_{\rm L} > \tau_{\rm R}$.  Comparing the magnitude of
these two currents we quantify the thermal rectification as
\begin{equ} \label{eq:rect}
\Delta = \frac{\max\{|J^+|,|J^-|\}}{\min\{|J^+|,|J^-|\}} ~.
\end{equ}

From our argument it is clear that rectification will be effective if one of
the temperatures is very small, say $\tau_{\rm L} \ll 1$ and the other is
simply above the critical, $\tau_{\rm R} > 1$.  It is possible to make a
quantitative prediction on the scaling of rectification factor $\Delta$ with
temperatures.  When the energy current through the magnetic field interface is
very weak, namely if $\tau_{\rm L} < \tau_{\rm R}$, the current is proportional
to the transmission coefficient at the interface, i.e.  one minus the
probability of reflection.

From our previous discussion it is clear that if $\tau_{\rm L} < 1 <
\tau_{\rm R}$ the particle density will be larger at the left  cell (with zero
magnetic field). This is mainly because the cold particles in the left
cell spend long time before being able to cross the interface. A particle of
velocity $v$ is transmitted (not reflected) to the right cell if it crosses
the interface at a distance from the upper boundary shorter than $2\rho(v)$
\cite{sign}.  Therefore, invoking the ergodicity of the dynamics in the left
cell, we can simply estimate the transmission coefficient $t$ as
\begin{equ} \label{eq:t}
t^+ \sim \frac{2\rho(v)}{\lambda} ~,
\end{equ}
where $\rho(v(\tau))=\sqrt{2mk_B\tau_{\rm min}}/eB$ and we denote by
$\tau_{\rm min} = {\rm min}\{ \tau_{\rm L},\tau_{\rm R}\}$.

However, in  the reverse situation  (exchanging $\tau_{\rm L}$  and $\tau_{\rm
  R}$) we have $t^- \sim 1$, so we can estimate the rectification
\begin{equ} \label{eq:scaling}
\Delta = t^-/t^+ \propto \frac{1}{\sqrt{\tau_{\rm min}}} ~.
\end{equ}

\begin{figure}[!t]
\begin{center}
\includegraphics[scale=0.7]{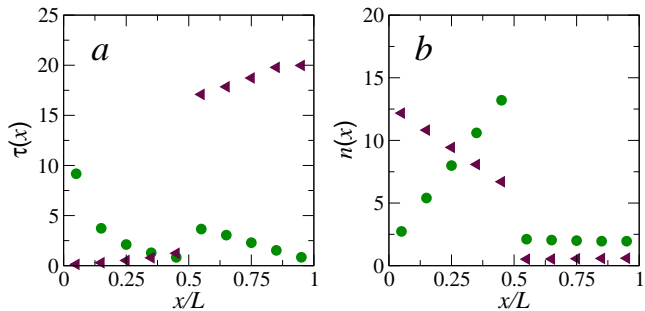}
\caption{
  Energy density ($a$) and particle density ($b$) profiles for a $10$-cell
  channel in a {\em step} configuration. The bath temperatures are 
  $\tau_\mathrm{min}=3.34277\times10^{-2}$,
  $\tau_\mathrm{max}=33.4277$.  In both panels profiles are shown for the
  positive and negative gradient for which $\tau_\R=\tau_\mathrm{max}$
  (triangles) and $\tau_\R=\tau_\mathrm{min}$ (circles) respectively.
\label{fig:T}}
\end{center}
\end{figure}

In Fig.~\ref{fig:rect-2} we show the rectification index $\Delta$ as a
function of $\tau_{\rm min}$ for fixed value of the maximal temperature
$\tau_{\rm max}$. We clearly confirm the scaling (\ref{eq:scaling}),
indicating also that the rectification index is only very weakly depending on
the maximal temperature (as long as $\tau_{\rm max} > 1$.  This can be seen in
the inset of Fig.~\ref{fig:rect-2}) where the rectification index $\Delta$ is
shown as a function of the two temperatures of the heat baths $\tau_\L$ and
$\tau_\R$. The correctness of the scaling \eref{eq:scaling} shows that the
magnetically induced rectification power is arbitrarily large for sufficiently
small temperature $\tau_{\rm min}$.

Even though a non-interacting system cannot reach local thermal equilibrium,
and therefore the concept of (local) temperature is not well defined, it is
instructive to compute temperature and density profiles, as long-time averages
of kinetic energy density and particle density as a function of the horizontal
coordinate along the chain.  In Fig.~\ref{fig:T} we plot the kinetic energy
density and the particle density measured in each cell of a channel of $10$
cells, the right half of $5$ cells being in a uniform magnetic field. We plot
the profiles for the positive gradient, $\tau_\L < \tau_\R$ (triangles), and
the inverted negative gradient (circles).  The positive gradient for which
$\tau_\L$ is $\tau_{\rm min}$, corresponds to the insulating case, \emph{i.e.}
to the situation in which the heat current is very small. In Fig.~\ref{fig:T}
we see that the insulating case corresponds to a large gap in the kinetic
energy profile. In contrast, for the negative gradient the gap in the kinetic
energy profile is much smaller and this coincides with the observation of a
larger heat current. However, note that the energy profile for the negative
gradient is not a monotonous function due to the lack of local thermal
equilibrium. Moreover, the density profile confirms our prediction that the
low current situation is characterized by a very small density of particles
on the side of the magnetic field.

\begin{figure}[!t]
\begin{center}
\includegraphics[scale=0.7]{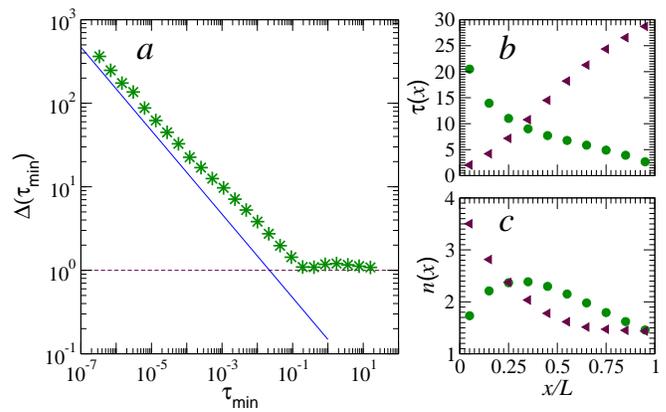}
\caption{
  $10$-cell channel in a \emph{gradient} configuration: ($a$) Rectification
  index $\Delta$ as a function of the minimal temperature $\tau_{\min}$, with
  $\tau_{\max}=33.4275$.  The temperatures are given in units of
  $T_\mathrm{c}(B)$, where $B=100$ is the strength of the magnetic field at
  the rightmost cell.  The dashed line is for the no rectification value
  $\Delta=1$.  The solid line corresponds to $\tau_{\min}^{-1/2}$.  In the
  panels at the right the profiles of kinetic energy ($b$) and particle
  density ($c$) for the positive (triangles) and negative (circles) gradient
  are shown.
\label{fig:grad}}
\end{center}
\end{figure}

We discuss now a slightly different situation in which the magnetic field does
not change abruptly from one half of the system to the other, but instead
changes gradually, forming a uniform gradient of the magnetic field. We refer
to this situation as the \emph{gradient} configuration for which the magnetic
field in each cell has an intensity given by $B_i=B(i-1)/(N-1)$ for
$i=1,2,\ldots,N$. Note that for $N=2$ the {\em gradient} and the {\em step}
configurations coincide.

\begin{figure}[!t]
\begin{center}
\includegraphics[scale=0.7]{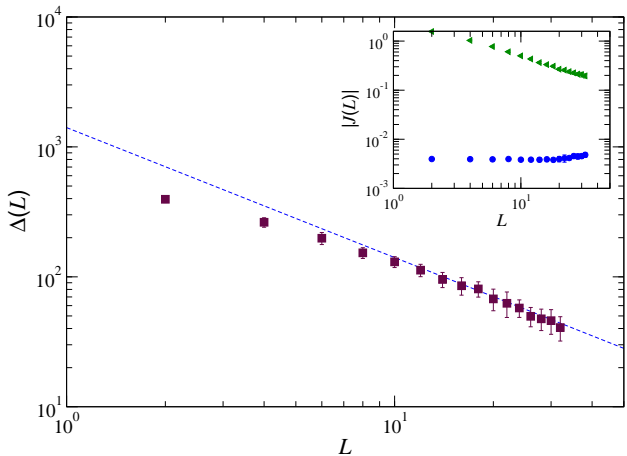}
\caption{
  Dependence of the thermal rectification index $\Delta$ as a function of the
  size of the chain $L$ for a chain in a \emph{step}-configuration with fixed
  $\tau_{\min}=3.34275\times10^{-3}$, $\tau_{\max}=3.34275\times10^{2}$ and
  fixed particle density, namely one particle per cell.  The dashed line corresponds to $L^{-1}$. Inset:
  heat current $J$ as a function of the size of the chain $L$ for the
  positive (triangles) and negative (circles) gradient.
\label{fig:vsL-step}}
\end{center}
\end{figure}

We have performed numerical simulations for the {\em gradient} configuration,
and we found no qualitative differences with respect to the {\em step}
configuration.  In Fig.~\ref{fig:grad} we show the dependence of the rectification
index on the minimal temperature (panel $a$).  As for the {\em step}
configuration we have found that for sufficiently low temperature $\tau_{\rm
  min}$, the rectification index again grows as $\Delta\sim1/\sqrt{\tau_{\rm min}}$.
In the panels ($b$) and ($c$) we show the profiles of the kinetic energy and
particle density respectively.  As expected the profiles for the {\em
  gradient} configuration are more smooth than for the {\em step}
configuration.

From our  analysis it  follows that our  rectification effect is  a
phenomenon which exists  only in  far from equilibrium  situation.
In  the thermodynamic limit,  a vanishingly  small temperature
gradient is  established  across the system.
In Fig.\ref{fig:vsL-step} we show the results of numerical
simulations  which indicate  that, for  fixed bath temperatures,
the rectification  index   scales  as  $\Delta   \sim 1/L$. In the
inset  of Fig.\ref{fig:vsL-step} the  time averaged heat currents
for the  positive and negative gradients are also shown.

Finally, it is interesting to give a quantitative estimate of the critical
temperature in Eq.~\eref{eq:Tc} in physical units.  Let us suppose that the gas of particles
inside the billiard of Fig.~\ref{fig:channel} consists of electrons.
Assuming that the dimension of the opening is $\lambda = 100 {\rm nm}$ and
a magnetic field  of $B = 1 {\rm T}$ is applied, then the critical temperature is
 $T_c\sim0.5 {\rm K}$.  Thus, a rectification power of $\Delta\sim10$ would
be measurable for thermal gradient given by $T_{\rm min}\sim10^{-3} {\rm
  K}$ and $T_{\rm max}\sim10 {\rm K}$ that appears accessible to nowadays
experiments.

In this letter we have presented  a novel mechanism for thermal rectification. 
This mechanism is fairly simple.   It results from the asymmetric behavior of
the dynamics  at the  magnetic interface, that  leads to a  simple temperature
dependence  of the  transmission  coefficient.  Moreover,  this mechanism  for
thermal  rectification  is not  based  on  the, macroscopically  unaccessible,
microscopic  particle interaction,  but on  the interaction  with  an external
field, making possible an easy control  of the power of rectification. We have
shown  that   the  thermal  rectification  power  is   arbitrarily  large  for
sufficiently small temperature of one of the heat baths.

Furthermore,  the physical  scales needed  for optimal  implementation  of our
theoretical  model  are  realizable  in present  nano-scale  experiments  with
meso-scopic  devices.   We  believe  that  it is precisely  at  these  scales  of
meso-scopic  physics where  such a  thermal rectifier  would  find interesting
applications.   This  fact makes  an  experimental  verification  of  the  mechanism
presented here very desirable.

\begin{acknowledgments}
CM-M acknowledge support by the Institute for Scientific Interchange.
TP acknowledges support by the program P1-0044 and project J1-7347
of Slovenian Research Agency.
\end{acknowledgments}


\begin{thebibliography}{}
\bibitem{sign} The  particles that  are transmitted to  the right  cell bounce
  repeatedly with either  the upper or the lower boundary  of the billiard and
  this depends on the sign of the magnetic field.
\end{thebibliography}
\end{document}